\documentstyle[12pt]{article}
\input math_macros.tex

\def\ref#1{$^{#1)}$}
\begin{document}
\begin{titlepage}
\begin{center}
        \hfill LBL-35864, July, 1994\\
         \hfill hep-ph/9407284\\

\vskip .05in

{\large \bf Anomalous Generation Numbers in SO(10) and Supersymmetric
SO(10) Unification Theories}

\vskip .05in
Huazhong Zhang\footnote{Email: Zhang@theorm.lbl.gov, Zhanghz@sscvx1.ssc.gov.
Permanent address: P. O. Box 17660, Jackson, MS 39217, USA}\\[.1in]

{\em  Theoretical Physics Group, MS 50A-3115, Lawrence Berkeley
Laboratory, University of California, Berkeley, CA 94720, USA}
\end{center}
\begin{abstract}
One of the interesting features in unification models and supersymmetric
unification models is that the chiral states of quarks and leptons in a family
including a right-handed neutrino can be fitted neatly into a fundamental
spinor representation (f.s) of dimension 16 for the SO(10) gauge group.
However,
it is shown in this paper that such a fundamental spinor representation of
SO(10) for Weyl fermions will generate global (non-perturbative) gauge
anomalies (of new type) when restricting to the $SU(2)\otimes SU(2)\otimes
SU(2)
\otimes SU(2)$ gauge subgroup. Such an example is the four SU(2) factors
obtained through the reduction of the subgroup $SU(2)\otimes SO(7)$ of SO(10)
with the SO(7) to the three SU(2) factors. The branching rule in this case is
given by $(f.s)\rightarrow (2-1-2-2)\oplus (2-2-1-2)$ in terms of dimensions. A
consistent gauge theory implies the gauge symmetry in a gauge subgroup, and
then needs to be well-defined when restricting to the gauge subgroup.
Consequently, a consistent SO(10) quantum theory needs to satisfy our selection
rule $N_f+N_{mf}=even\geq 4$, namely the total number of generations with $N_f$
ordinary fermion families and $N_{mf}$ mirror fermion families is even and
larger than three, and the three generations of chiral fermions in this content
can not correspond to a consistent theory. Then we expect that there exist at
least one additional fermion family including a right-handed neutrino or at
least one family of mirror fermions including a left-handed mirror neutrino
if SO(10) unification theory is relevant to our realistic world. The next
possibility is to have total six generations that may subject to stringent
experimental constraints. Two such more plausible examples are obviously
(i) there are three more generations of ordinary fermions including
three right-handed neutrinos correspondingly, or
(ii) there are three more generations of mirror fermions including three
left-handed mirror neutrinos correspondingly. A brief sketch of some
physical implications are also given. In the case of mirror fermions,
a fundamental aspect is that there will be also V+A currents coupling to the W
bosons and the weak interaction is not completely chiral. We also noted
the absence of global anomaly in gauge subgroups of the gauge group
SU(5),$E_6,~ E_8$, SU(4), SU(6), etc. for the relevant representations.

\end{abstract}
\end{titlepage}
\newpage
\renewcommand{\thepage}{\arabic{page}}
\setcounter{page}{1}
Since the discovery of Yang-Mills theories$^1$, elementary particle
physics has gained great development in the framework of non-abelian
gauge theories. With the construction of standard
electroweak gauge theory$^2$, one of the most interesting ideas has
been incorporating the standard model into a grand unified theory$^{3-4}$
(GUT) or a supersymmetric grand unified theory$^{5-6}$. It is now known
that although the minimal SU(5) model$^2$ is not compatible with proton
decay search$^7$ and CERN LEP data$^8$, the unification may be achieved
in either supersymmetric GUT gauge theories or a GUT with a gauge group
larger than SU(5) spontaneously broken to the standard gauge group in
at least two stages. Such an example$^9$ is the SO(10) GUT model which
may break$^{10}$ first to left-right symmetric
$SU(3)\otimes SU(2)_L\otimes SU(2)_R\otimes U_{B-L}$ model at a scale $M_X$
and then to the standard model. While the consideration of gauge
hierarchy certainly prefers a supersymmetric SO(10) theory, without
evidence of supersymmetry at low energies, non-supersymmetric SO(10)
models are still of interest$^{10}$. Among many attractive physics
features$^{11}$ of SO(10) models are that they preserve a good prediction for
the $\sin^2{\theta}_w$, permit neutrinos having small masses through see-saw
mechanism, and may give predictions for some parameters in the standard
model. Another interesting feature in these models is the
structure of the group representation itself, which has all the 16
chiral states of quarks and leptons including the right-handed neutrino
that comprises one family fitting neatly into a fundamental spinor
representation of dimension 16 for the SO(10) unification gauge group.
This is the feature we will focus on in this paper. Especially, we will
analyze the global (non-perturbative) effects of this group theoretic
structure. Then, the relevant remarks and consequences will be discussed.
As a matter of fact, we will see that the SO(10) models with Weyl
fermions in a fundamental spinor representation will generate a new type of
global (non-perturbative) gauge anomalies$^{12}$ which will be clarified soon,
and therefore, the corresponding generating functional is ill-defined, or the
quantum theory is inconsistent. This also suggests that the possibilities for
this type of global gauge anomalies need to be taken into consideration
carefully in general in non-abelian gauge theories with Weyl fermions.

It was shown by Witten$^{13}$ that an SU(2) gauge theory in four dimensions
with an old number of Weyl fermion doublets is mathematically inconsistent
due to a global (non-perturbative) gauge anomaly in the theory. With a global
gauge anomaly, the generating functional for the quantum theory is ill-defined.
Topologically, this is associated with the fact that the homotopy group
$\Pi_4(SU(2))=Z_2$ is non-trivial. In general, for a non-abelian gauge theory
in an even dimensions D=2n, one still needs to consider the possibility of
gauge anomalies if the homotopy group $\Pi_{2n}(G)$ for the gauge group G
is non-trivial when the theory is free of local (perturbative) gauge
anomalies. Such possible global gauge anomalies have been investigated
for SU(N) gauge groups$^{14-21}$, and systematically and rather generally for
arbitrary compact and connected simple gauge groups in generic even
dimensions$^{15-21}$, especially in terms of$^{15-21}$ the James numbers of
Steifel manifolds and generalized Dynkin indices. In this paper, we have shown
that Weyl fermions in a fundamental spinor representation of SO(10) will
generate global gauge anomalies in the theory. Therefore, the conventional
SO(10) and supersymmetric SO(10) unification models must be modified in order
for the theory to be free of the global gauge anomalies.

     The new type global gauge anomalies relevant here are topologically
more subtle than the usual ones similar to that first noted by Witten$^{13}$
with the non-trivial homotopy group$^{22}$ $\Pi_4(G)$ for the gauge group G
in four dimensions, since the relevant homotopy $\Pi_4(SO(10))={0}$ is trivial.
Therefore, one might usually expect that there should not be
any global gauge anomalies, where note that the
SO(10) gauge theories are free of local anomalies in four dimensions in
which it is well known that$^{23}$ local gauge anomalies
can only arise from Weyl fermions in the complex representations of
SU(N) ($N\ge 3$). Even for a simple subgroup H of a gauge group G with
$\Pi_{2n}(G)={0}$ being trivial, we showed the following proposition$^{16}$.
{\em Proposition}: In arbitrary D=2n dimensions, if the relevant Weyl
fermion representation in G free of local gauge anomaly in the strong anomaly
cancellation condition $TrX^{n+1}=0$ reduces to an irreducible representation
$\omega$ plus H (simple) singlets (in the generalized convention of allowing
negative multiplicities$^{15-16}$), then there will be no H global gauge
anomalies for the representation $\omega$. For the special case with a SU(2)
group embedded into a simple gauge group G with $\Pi_{4}(G)={0}$,
it is known that there will be no SU(2) global gauge anomaly if the
local anomaly-free condition for the group G is satisfied$^{24}$.

The new global gauge anomalies we note arises from the restriction
of a gauge group G with relevant trivial homotopy group to its
gauge subgroups (containing more than one simple ideals with non-trivial
forth homotopy group). The example we will
focus in this paper is the subgroups of SO(10) due to its crucial importance
and relevance to the unification theories.
We will now describe our result for the SO(10) gauge theories, and may use the
Lie algebras for the discussion of representations.
The same notations may be used for the Lie groups and corresponding
Lie algebras, no confusion should be caused in our discussion here.
The SO(10) group contains a maximal subgroup $SU(2)\otimes SO(7)$
(there is not difference between SU(2) and SO(3) in our consideration since
they have the same forth homotopy group). Restricting the SO(7) to the
subgroup $SU(2)\otimes SU(2)\otimes SU(2)$, then we obtain a
subgroup $SU(2)\otimes SU(2)\otimes SU(2)\otimes SU(2)$.
For this subgroup, we have the relevant
homotopy group
\begin{equation}
\Pi_{4}(SU(2)\otimes SU(2)\otimes SU(2)\otimes SU(2))
=Z_2\oplus Z_2\oplus Z_2\oplus Z_2.
\end{equation}
The homotopy group topologically classifies the continuous gauge
transformations restricted to this subgroup in the compactified spacetime
manifold. The non-trivial topological $Z_2\oplus Z_2\oplus Z_2\oplus Z_2$
structures exist when the SO(10) gauge theory is restricted
to the subgroup. The possible global gauge anomalies for a SU(2)
gauge theory in arbitrary 2n dimensions have been determined
completely$^{16}$ in terms of James numbers of Stiefel manifold
(see ref.16 and references therein) and the generalized
Dynkin indices$^{25-26}$.
The branching rule$^{27}$ for a fundamental spinor representation
(f.s) of SO(10) in such a reduction is given by
\begin{equation}
(f.s)\rightarrow (1-0-1-1)\oplus (1-1-0-1)
\end{equation}
in terms of Dynkin labels or
\begin{equation}
(f.s)\rightarrow (2-1-2-2)\oplus (2-2-1-2)
\end{equation}
in terms of dimensions.

To determine the possible global anomalies, we will first consider an
irreducible representation (2-1-2-2) in the branching rule eq.(3), then
the overall possibilities can be clarified. For the irreducible
representation (2-1-2-2), it is equivalent to consider the possible global
anomaly for the group as three SU(2) factors in the irreducible representation
(2-2-2). We can embed this into the SU(8) in the fundamental representation
$\Box$. Through the reduction of SU(8) to $SU(2)\otimes SU(4)$, and then
with the SU(4) to two SU(2) factors, we obtain the three SU(2) factors
in the irreducible representation (2-2-2). Since $\Pi_4(SU(8))$ is trivial,
the embedding condition is satisfied, in other words, the possible
$SU(2)\otimes SU(2)\otimes SU(2)$ global gauge anomalies will appear as
the Wess-Zumino term for the SU(8) in the fundamental representation.
With the fact $\Pi_5(SU(8))=Z$, and using exact homotopy sequence
it is known$^{15-20}$ that the basic global anomaly coefficient is this case
is given by $A=exp\{i\pi Q_2(\Box)\}$. The $Q_2(\Box)$ is the
second-order Dynkin index for the $\Box$ of SU(8) with the possible constraint
that the SU(8) gauge theory should be free of local anomaly when restricting to
the three SU(2) factors on the spacetime $S^4$ as the boundary of a
five-dimensional disc $D^5$. However, since the three SU(2) factors are
automatically free of local gauge anomaly in four dimensions, there is no
constraint on the Dynkin index. Therefore$^{25-26}$, we have $Q_2(\Box)=1$,
and the basic global anomaly coefficient in this case is then $A=-1$. We note
that$^{15-16}$ in order to obtain the topologically non-trivial SU(8)
transformation in the $\Pi_5(SU(8))=Z$, the corresponding gauge transformation
on the $D^5$ needs to be topologically non-trivial in all the three SU(2)
factors when restricting to its boundary $S^4$. Only in
this case, none of the SU(2) factors may be topologically reduced or
equivalently to give a factor 2 from the dimension as a multiplicity for the
other SU(2) factors, so that the eight-dimensional (irreducible) embedding is
topologically effective.
Obviously, we have determined that the basic global anomaly coefficient for
the the representation in eq.(3) for the four SU(2) factors is given by
\begin{equation}
A=(-1)\otimes (-1),
\end{equation}
and the theory with such a restriction has $Z_2\oplus Z_2$ global gauge
anomalies. The consequence is that$^{13}$ the generating functional and
the operators invariant under such gauge subgroup cannot be well-defined
relative to the relevant large and continuous gauge transformations
in the subgroup. The $Z_2\oplus Z_2$ anomaly may be understood
as that when the gauge transformation is topologically non-trivial in three of
the SU(2) factors simultaneously but trivial in either the second or the third
one in our notation, the fermion measure will change a sign, the quantum theory
is then not well-defined$^{13}$. Note that it is necessary for the gauge
transformations to be topologically non-trivial in the first and forth
SU(2) factors also to generate a global anomaly, since only in this case,
they cannot be continuously deformed into identity transformation in these two
SU(2) sectors and they will not contribute a factor 2 from each of the
dimensions for the two SU(2) factors.

{\em Remark}: The two different $Z_2$ global gauge anomalies for the
$Z_2\oplus Z_2$ arise from the two different irreducible representations
in the branching rule eq.(3) and correspond to topologically inequivalent
gauge transformations. Therefore, the fact that the SO(10) gauge group with
$\Pi_4(SO(10))=\{0\}$ does not have local gauge anomalies will not contradict
to our result. This is because for each of the irreducible representations
in eq.(3) (e.g the (2,1,2,2) for the four SU(2) factors), it cannot be
embedded into a representation $\omega$ of SO(10) such that the $\omega$
reduces to the irreducible representation ((2,1,2,2)) plus singlets upon the
reduction. This is also an explicit example showing that the conventional
proposition$^{16}$ noted by using the Wess-Zumino term argument does not apply
generally to the case in which the relevant subgroup has more than one
ideals with non-trivial 2n-th homotopy group in D=2n dimensions and the
representation is not irreducible.
This is also why a fundamental representation (f.s) of the SO(10) cannot have
$SP(4)\otimes SP(4)$ global anomaly, due to the fact that the (f.s) reduces
to the sixteen-dimensional {\em irreducible} representation ($\Box - \Box $)
upon the reduction $SO(10)\downarrow SP(4)\otimes SP(4)$, the conventional
argument of using Wess-Zumino term for the SO(10) may apply. In this case,
the vanishing of SO(10) local anomaly or Wess-Zumino term implies the
absence of the relevant $SP(4)\otimes SP(4)$ global anomaly.
The problem with embedding a direct sum with more than
one irreducible representations of global gauge
anomalies is that the anomaly information may not be extracted independently
due to the fact that$^{15}$ the global gauge anomaly for an irreducible
representation free of local gauge anomaly can be at most of $Z_2$ type.
Generally, from this point of view for the global gauge anomalies, the
restriction of a gauge theory to a gauge subgroup H may not be the same as
embedding the gauge subgroup H into the original gauge group G due to the
representation condition needed to analyze the possible global gauge anomalies.

Note that generally gauge symmetry in a gauge group implies the gauge symmetry
in its gauge subgroup (see ref.28 for the other studies related to this
property), namely a well-defined gauge theory needs to be well-defined when
restricting to its gauge subgroups. In quantum theory,
if there are gauge anomalies when restricting to a gauge subgroup,
then the gauge theory cannot be well-defined. In conclusion, the SO(10)
gauge theories with Weyl fermions in a fundamental spinor representation
of dimension 16 have global (non-perturbative) gauge anomalies.
The SO(10) has two fundamental spinor representations which are
complex conjugate to each other. Our results applies to
either one of them.

Denote the numbers of two inequivalent fundamental spinor representations
as $N(16)$ and $N(\bar{16})$, obviously we need to have
\begin{equation}
N(16)+N(\bar{16})=even,
\end{equation}
in order to cancel out the global gauge anomalies.
Consequently, SO(10) unification models
with three generations of fermions have global gauge anomalies.
We have checked that the adjoint representation of dimensions 45 for
the SO(10) will be free of global gauge anomalies for the relevant subgroups.
Therefore, our conclusion applies both to the non-supersymmetric SO(10) models
and supersymmetric models in which gauginos are in the adjoint representation.

The physics consequences of our result may be of fundamental interest if
the SO(10) gauge theories are relevant to the realistic world.
Obviously, the SO(10) models and supersymmetric SO(10) unification models need
to be modified according to our analysis. In the usual physics
convention for the Weyl fermions with the observed three families of leptons
and quarks, our selection rule is written as
\begin{equation}
N_f+N_{mf}=even\geq 4,
\end{equation}
with the $N_f=N(16)$ and $N_{mf}=N(\bar{16})$ denoting the number of
fermion families and the number of mirror fermion families respectively.
The above equation is our main result in this paper.
Therefore, we predict that there will be at least one more fermion family
or at least one mirror fermion family if an SO(10) unification
gauge theory is realistic. Where in the content of SO(10) unification,
the fourth generation (or a generation of mirror fermions) also includes
a right-handed neutrino (or a left-handed mirror neutrino).
Mirror fermions have the
same $SU(3)\otimes SU_L(2)\otimes U(1)_Y$ quantum numbers as the ordinary
fermions except that they have opposite handedness. Usually$^{29}$,
mirror fermions are considered with three generations. Conventionally, one
family of mirror fermions seems not so motivated. However, our result of the
global gauge anomalies shows that it is one of the simple ways to cancel the
global anomalies. As in the usual discussions, if there exists fourth
generation of fermions with V-A weak interaction, then of course, it seems
natural to have either no mirror fermions or four families of mirror fermions.
{}From the anomaly-free point of view. The next possibility is either to
have three generations of ordinary fermions and three generations of mirror
fermions correspondingly as in the usual discussions of mirror fermions,
or to have six generations of fermions with three more repetitions of an
ordinary fermion family. If there are mirror fermions, one of the
most fundamental consequences will then be that the Lorentz structure
of the weak interaction will no longer be chiral with only V-A currents
coupling to the W gauge bosons, there will be also
V+A piece which though may be very small relevant to the current
experimental observation$^{29}$. There has been analysis$^{29}$ about the
charged and neutral current data suggesting that the possible V+A
impurity in the weak amplitudes is typically less that about 10\%.

We will now give a brief sketch of some other related physics issues,
for details see the relevant references.
In the content of the electroweak theory, either an additional
generation of fermions or a generation (three generations) of mirror fermions
obtain their masses through the electroweak symmetry breaking at the order of
about O(300Gev), this will give effects on low energy physics and also subject
to both theoretical and experimental constraints. The LEP date set a lower
bound for their masses denoted by $M_F$ at about $M_F\geq m_z/2$, namely
about half of the Z boson mass. The partial wave unitarity$^{30}$ at high
energies shows that the masses above about
$O(600Gev$/$\sqrt{N_{DQ}})$ and $O(1Tev$/$\sqrt{N_{DL}})$ for quarks
(or mirror quarks) and leptons (mirror leptons) will signal the breakdown of
the perturbation theory, where $N_{DQ}$ denotes the
total number of nearly degenerate weak-isospin doublets for quarks and
mirror quarks, and similarly with $N_{DL}$ for leptons and mirror leptons.
There may be stringent constraint on the masses, mass splittings in a
weak-isospin doublet for possible new fermions due to the bound on the
correction $\delta\rho$ for the parameter$^{31,32}$
$\rho={m_w}^2$/${m_z}^2\cos^{2}\theta_{w}$
from its tree level value in the minimal standard model,
as well as for the other precision electroweak parameters$^{31, 33}$.
It is known that the radiative corrections$^{31}$
in perturbation theory can play an active role in this. There are also
recent discussions that$^{32}$ the possible bound states formed by the exchange
of Higgs bosons in the presence of additional heavy fermions may give
non-perturbative contribution $\delta\rho<0$ and cancel the perturbative
correction within the current experimental error in the nearly degenerate case,
and therefore can relax the constraints for the masses and mass splittings
due to the $\rho$ parameter.

An additional family of fermions or mirror fermions may also be constrained by
that Yukawa couplings should remain small during the evolution
in the perturbative region
(about $\alpha_{Yuk}={\lambda_{Yuk}}^2$/$4\pi\leq 1$), otherwise its
running may induce Landau poles in the one-loop approximation. The presence
of these singularities at some scale signals the breakdown of perturbation
theory and the probable triviality of the continuum limit. Related to
the running of Yukawa couplings and infrared fixed-point solution$^{34}$ to
the renormalization group equations, there has been discussions$^{34, 35}$
in supersymmetric unification models with Yukawa coupling unification
(e.g. $\lambda_{\tau}=\lambda_b$ at the unification scale) that only
small regions in the $m_t-\tan\beta$ ($\tan\beta=v_{up}$/$v_{down}$) plane may
be allowed (e.g. about $1\leq\tan\beta\leq 1.5$ or $\tan\beta\geq 40\pm 10$
with $m_t\leq 175 Gev$). The value of $\tan\beta$ can typically
effect$^{34-35}$ the flavor changing neutral currents in processes like
$b\rightarrow s\gamma$ and $B\bar{B}$ mixing and to the proton decay$^{36}$.
It also constraints on the Higgs boson masses$^{37}$ which may be
relevant to the LEP II. If there are additional fermions or mirror fermions,
one may expect that their presence will also have these typical effects,
relevant discussions with Yukawa coupling unification at the unification scale
then may need to incorporate them. At least, the possibility of
one additional generation of mirror fermions from our motivation in terms
of global gauge anomalies sounds quite new.
It has been argued$^{38}$ in the conventional mirror fermion models (with
three generations of mirror fermions corresponding to ordinary observed
generations of quarks and leptons) that mirror doublets should always be
assumed degenerate in masses (without considering the non-perturbative effects
in ref. 32) in order to reproduce the precision LEP data, and the possible
Higgs masses may be in rather restricted regions.
We note that mirror fermions of at least three generations
usually appear in the particle spectrum of many theories other than some
superstring models with family unification, such as those with extended
supersymmetry ($N\geq 2$) imposed on a gauge theory, in the
Kaluza-Klein theories$^{39}$, and some composite models$^{40}$.
However, according to our analysis of global
anomalies in SO(10) unification gauge theories, one of the interesting models
is to have only one generation of mirror fermions besides the three
generations of ordinary fermions. In general, one may expect that$^{29,38}$
mirror fermions need to mix with the ordinary fermions in order to avoid
stable mirror fermions although the mixing may be small.
If there exists only one generation of mirror fermions,
fundamentally, it is unnatural to assume that it corresponds to a particular
family of ordinary fermions. Therefore, this generation of mirror fermions
will mix with all the three generations of ordinary fermions, and this then
will induce the flavor mixing between the three generations of ordinary
fermions also, this seems to provide another origin for the possible flavor
mixing and possible CP violations. Moreover, as it is known that
the lifetime of heavy neutrinos may subject to cosmological constraint$^{41}$
(a suggestion is about $\tau < 10^3 yr (1kev/m_v)^2$) since it may be strongly
believed that the age of the universe is greater than $10^{10}$ years.
Our selection rule may have consequences on the structure of the universe.

According to our rule in eq.(6) for SO(10) unification gauge theories, since
the total number $N_t$ of generations for ordinary fermions and mirror fermions
needs to be $N_t=even\geq 4$, the models with family unification beyond SO(10)
theories with the other possibilities we discussed above other than the case
$N_f=N_{mf}=3$ are then better motivated. The models with only $N_t=N_f=3$ or
$N_t=N_f+N_{mf}=odd$ when the gauge symmetry is broken down to the SO(10)
then may not be consistent. Our rule for the
generation numbers in eq.(6) may be of fundamental interest and importance.
We will conclude our paper with the following related discussions.

(i)We have also noted that the $SU(2)\otimes SU(2)\otimes SU(2)\otimes SU(2)$
is the only possible subgroup up to isomorphism having $Z_2\oplus Z_2$
global gauge anomalies.
Another example of the reduction is the four SU(2) factors obtained
through the reduction of the subgroup $SU(2)\otimes SU(2)\otimes SO(6)$
with the SO(6) to two SU(2) factors. But there may be possibly
$Z_2\oplus Z_2\oplus Z_2\oplus Z_2$ anomaly in
the four SU(2) factors obtained through the reduction of the subgroup
$SP(4)\otimes SP(4)$ of SO(10) with each SP(4) to two SU(2) factors,
since this reduction leads to a different branching rule.
Moreover, the SO(10) group in low-dimensional representations of
dimensions 45, 54, 120 etc. will not have global gauge anomalies.
Especially, we emphasize that since in the supersymmetric SO(10) models,
the gauginos are in the 45-dimensional adjoint representation,
our selection rule eq.(6) applies both to non-supersymmetric SO(10) and
supersymmetric SO(10) unification theories.

(ii)For Weyl fermions in the $(\Box - \Box)$ for the $SU(2)\otimes SU(2)$,
it can be embedded into the four-dimensional $\Box $ of SU(4) with
$\Pi_5(SU(4))=Z$. One can easily see that there may be possibly
a $Z_2$ global gauge anomaly in this case. This fact is relevant to
the possible $Z_2\oplus Z_2\oplus Z_2\oplus Z_2$ global anomaly in (i)
for a (f.s) of SO(10).
By using this fact, one can also check that the vector representation of
dimension 10 for the SO(10) may have global gauge anomalies when
restricting to certain subgroups as product of SU(2) factors through many
reductions.
But they all cancel out if the total number of the vector representations are
even. Some examples are given below.
(1) $Z_2\oplus Z_2\oplus Z_2\oplus Z_2$
global anomalies when restricting to the
$SU(2)\otimes SU(2)\otimes SU(2)\otimes SU(2)$
through the reduction of subgroup $SP(4)\otimes SP(4)$ of SO(10);
(2) $Z_2\oplus Z_2$ global gauge anomalies when restricting to a
$SU(2)\otimes SU(2)$ gauge subgroup through the following reductions:
(2a)The $SU(2)\otimes SU(2)$ from the reduction of either one of
the SP(4) factors in (1);
(2b) the $SU(2)\otimes SU(2)\otimes SO(6)$ ($SO(6)\cong SU(4)$)
subgroup of SO(10) to the two SU(2) factors;
(2c) by the reduction of the subgroup $SU(2)\otimes SO(7)$ of SO(10) to
$SU(2)\otimes SU(2)\otimes SU(2)\otimes SU(2)$, the
relevant two SU(2) factors both are in the fundamental representation
in one of the irreducible representations (2-2-1) in the branching
rule for the reduction of the SO(7) to $SU(2)\otimes SU(2)\otimes SU(2)$.
Obviously, the global gauge anomalies in a fundamental representation
cannot be canceled by adding more vector representations and vice versa.
It is a typical feature that the branching rules become much more involved
when the dimension goes higher, if there were other possible anomalies
they would be very dependent on the reduction procedure, global anomalies
from representations of different dimensions may not cancel each other.
At least, we checked up to dimensions at least several hundred, no other higher
irreducible representations can have the same possibilities for the global
anomalies as that of a fundamental spinor representation. Our selection
rule for the generation numbers in eq.(6) is realistically general.

(iii) For superstring theory with gauge group $E_8\times E_8'$, a
compactification of the heterotic string is $E_6\times E_8'$, N=1
supersymmetric Yang-Mills theory in four dimensions$^{42}$.
In the $E_6$ sector, the left-handed Weyl fermions
are in the real representation
$\{78\}\oplus 3\times\{27\}\oplus 3\times\{\bar{27}\})\oplus 8\times\{1\}$
of the $E_6$ in terms of the dimensions.
In four dimensions, the $E_6$ is a local anomaly-free group with
$\Pi_4(E_6)={0}$ being trivial. We can show that this representation
will not have a global anomaly when restricting to a subgroup with
non-trivial forth homotopy group.
In this case of $E_6$ upon the reduction to SO(10), it can also be seen more
obviously by our analysis. Upon the reduction
$E_6\downarrow SO(10)$, $78\rightarrow 45\oplus 16\oplus\bar{16}\oplus 1$,
and $27\rightarrow 16\oplus 10\oplus 1$ and correspondingly for the $\bar{27}$.
After the decomposition, there is a $\bar{16}$ for each 16 and the 10 also
appear in pairs. Therefore, there will be no global gauge anomalies upon
reduction to SO(10). One can also see explicitly that$^{12}$ the theory have no
global gauge anomaly for the other possible gauge subgroups
with non-trivial forth homotopy group. Therefore, the relevant heterotic
string theory is free of both local and global gauge anomalies. An $E_6$
unification gauge theory with even total number for 27 and its
complex conjugate $\bar{27}$ is free of global gauge anomaly.

(iv) For the SU(5) and supersymmetric SU(5) theories, we have shown that
the relevant Weyl fermion representations (e.g. $\bar{5}\oplus 10$)
free of local gauge anomaly are also free of global gauge anomaly
for the gauge subgroups (such as SP(4) and $SU(2)\times SU(2)$ etc.),
as well as$^{24, 16}$ SU(2) with non-trivial forth homotopy group.
Moreover, for the relevant representations of $E_8$, SU(4)
(for example $8\{4\}\oplus\bar{10}$), and SU(6) gauge groups free of
local gauge anomaly, it can be seen that$^{12}$ they are free of
global gauge anomaly for the other subgroups with non-trivial relevant
homotopy group.
The central result of the paper is our rule of selecting the even
total generation numbers for ordinary fermions and mirror fermions.
Explicit details of the other discussions above are too involved
to be given here.
Our analysis also suggests that a new direction of research is to
consider the possible gauge anomalies in the subgroups larger than
SU(2), especially for those containing more than one simple ideals
with non-trivial forth homotopy groups. A general and
useful result is that$^{16}$ in D=4 dimensions (mod 8), SU(2) has
$Z_2$-type global gauge anomalies only if an irreducible representation
has the spin of the form J=(1+4k)/2=1/2,5/2,9/2,....


The author would like to thank T. Han, Z. Huang, P. Q. Hung, H. Murayama and
D. D. Wu for valuable discussions. The author also like to thank
I. Hinchliffe and the theoretical Physics group at LBL for hospitality.


\begin{thebibliography}{99}
\bibitem{thfcnc}General consideration for other groups will be given elsewhere.
\end{thebibliography}
\end{document}